\documentclass[prl,twocolumn,showpacs,aps,floatfix,superscriptaddress]{revtex4}
\usepackage{graphicx}
\begin{document}

\title{Nonlocal effects on magnetism in the
diluted magnetic semiconductor Ga$_{1-x}$Mn$_{x}$As}

\author{Unjong Yu}
\affiliation{Department of Physics and Astronomy \&
Center for Computation and Technology,
Louisiana State University,
Baton Rouge, LA 70803, USA}
\affiliation{Department of Applied Physics,
Gwangju Institute of Science and Technology, Gwangju 500-712, Korea}
\author{Abdol-Madjid Nili}
\affiliation{Department of Physics and Astronomy \&
Center for Computation and Technology,
Louisiana State University,
Baton Rouge, LA 70803, USA}
\author{Karlis Mikelsons}
\affiliation{Department of Physics and Astronomy \&
Center for Computation and Technology,
Louisiana State University,
Baton Rouge, LA 70803, USA}
\affiliation{Department of Physics, University of Cincinnati, Cincinnati, Ohio 45221, USA}
\author{Brian Moritz}
\affiliation{
Stanford Institute for Materials and Energy Science,
SLAC National Accelerator Laboratory and Stanford University, Stanford, CA 94305, USA
}
\author{Juana Moreno}
\author{Mark Jarrell}
\affiliation{Department of Physics and Astronomy \&
Center for Computation and Technology,
Louisiana State University,
Baton Rouge, LA 70803, USA}

\date{\today}

\begin{abstract}
The magnetic properties of the diluted magnetic semiconductor
Ga$_{1-x}$Mn$_{x}$As are studied within the dynamical cluster approximation.
We use the $\mathbf{k}\cdot\mathbf{p}$ Hamiltonian to describe the 
electronic structure of GaAs with spin-orbit coupling and strain effects.
We show that nonlocal effects are essential for explaining the
experimentally observed transition temperature and saturation magnetization.
We also demonstrate that the cluster anisotropy is very strong and induces
rotational frustration and a cube-edge direction magnetic anisotropy at low temperature.
With this, we explain the temperature-driven spin reorientation in this system.
\end{abstract}


\maketitle


The discovery of high temperature ferromagnetism in diluted magnetic
semiconductors (DMS) has stimulated a great deal of attention \cite{Ohno98}.
The interest in these materials is due to possible applications in
spintronics \cite{spintronics} as the source of a spin polarized
current or as the base material for a chip that can simultaneously store
and process data.

In spite of extensive studies,
our understanding of ferromagnetism in these systems is far from complete \cite{DMS_theory}.
There are a few serious difficulties in the theoretical study of DMS:
(i) The magnetic interaction between local magnetic moments and itinerant carrier spin,
which is responsible for the high transition temperature ($T_c$),
is strong and outside the Ruderman-Kittel-Kasuya-Yosida (RKKY) regime.
(ii) There exists strong disorder from the random
distribution of magnetic ions. (iii) Nonlocal effects are expected to be
crucial judging from the spatially oscillating and anisotropic magnetic interaction
predicted in theory \cite{Zarand,Moreno06,Mahadevan,Kudrnovsky} and
observed in experiments \cite{Kitchen}.

The mean-field study by Dietl \textit{et al.} \cite{Dietl00} captures the main 
features of DMS systems qualitatively and some even quantitatively.
However, it ignores strong correlations, disorder effects, and spatial 
fluctuations, and fails to describe some DMS materials,
such that subsequent studies have brought their approach into
question \cite{Priour,Bouzerar}.  Studies \cite{DMS_DMFT} based on the dynamical 
mean-field theory (DMFT) \cite{DMFT} have made considerable improvements by 
including strong correlation and disorder effects.  However, the local 
nature of the DMFT presents severe limitations when studying this system.
The effects of short-range fluctuations and 
spatial correlations were shown to be important in the classical
Heisenberg model [14]. In this letter, we show that nonlocal effects
may be equally important for the itinerant carriers, which mediate
the effective interaction between local moments in DMS.

The dynamical cluster approximation (DCA) \cite{DCA} systematically incorporates
nonlocal effects as the cluster size ($N_c$) increases while retaining
strong correlations.
When $N_c=1$, the DCA is equivalent to the DMFT, and exact results are
approached as $N_c\rightarrow \infty$.
Since all the possible disorder configurations are considered in a cluster,
the DCA is also a better approximation for disorder average than the coherent potential 
approximation or DMFT by including multi-impurity scattering terms \cite{DCA_disorder}.
Thus, the DCA is an ideal method for studying DMS systems. In this Letter, we 
study the magnetic properties of the prototypical DMS system Ga$_{1-x}$Mn$_x$As 
using the DCA and the $\mathbf{k}\cdot\mathbf{p}$ method, which describes the 
non-interacting band structure of pure GaAs. We show that nonlocal effects 
are very important for properly capturing the magnitude of $T_c$, the saturation 
magnetization, and the magnetic anisotropy of this material.
In particular, we show that the strong cluster anisotropy is responsible for
the magnetic anisotropy along the cube-edge direction 
and the spin reorientation at low temperature.


The model Hamiltonian we adopt is 
\begin{eqnarray}
H=H_{\mathbf{k}\cdot\mathbf{p}} + J_c \sum_{I} \mathbf{S}(R_I) \cdot \mathbf{J}(R_I),
\label{Hamiltonian}
\end{eqnarray}
where the first term describes the electronic structure of the host material (GaAs) in the
$\mathbf{k}\cdot\mathbf{p}$ approximation and the second term introduces a magnetic interaction
between the carrier spin ($\mathbf{J}$) and the local magnetic moment ($\mathbf{S}$) of Mn 
at position $R_I$.  The large magnitude of the Mn magnetic moment ($S=5/2$) allows us to 
treat it classically.
This model is generally accepted to describe DMS 
\cite{Zarand,Moreno06,Dietl00,DMS_DMFT,Abolfath,Dietl01}, since
a mean-field treatment of the Hamiltonian~\cite{Dietl00,Abolfath,Dietl01} 
is able to explain many physical properties of the system.
For the $\mathbf{k}\cdot\mathbf{p}$ Hamiltonian 
($H_{\mathbf{k}\cdot\mathbf{p}}$), we adopt a 4$\times$4 Luttinger-Kohn model describing 
heavy and light hole bands with spin-orbit coupling, but ignoring the conduction and 
split-off bands.  We use the Luttinger parameters
$\gamma_1 = 6.98$, $\gamma_2 = 2.06$, and $\gamma_3 = 2.93$ \cite{Piprek}.
Biaxial strain is included in $H_{\mathbf{k}\cdot\mathbf{p}}$
through the strain tensor 
$\varepsilon_{xx} = \varepsilon_{yy} = \varepsilon_{0} = \Delta a / a$ and 
$\varepsilon_{zz} = (-2c_{12} / c_{11}) \varepsilon_{0}$
with the ratio of elastic stiffness constants $c_{12}/c_{11} = 0.46$.
Parameter $a$ is the lattice constant of
Ga$_{1-x}$Mn$_x$As, and $\Delta a$ is the difference between
the lattice constants of Ga$_{1-x}$Mn$_x$As and the substrate.
We use the hydrostatic deformation potential $a_v = 1.16$ eV and
the shear deformation potential $b=-2.0$ eV \cite{Piprek}.

In addition to the parameters of the $\mathbf{k}\cdot\mathbf{p}$ Hamiltonian,
we must determine the value of the exchange coupling $J_c$.  It can be obtained from
photoemission \cite{Okabayashi}, infrared \cite{Linnarsson,Singley}, and resonant
tunneling \cite{Ohno_Jc} spectroscopy and magneto-transport experiments \cite{Omiya},
which give $J_c$ = 0.6-1.5 eV. We adopt $J_c$ = 1 eV throughout this Letter.

\begin{figure}[tbp]
\includegraphics[width=8.2cm]{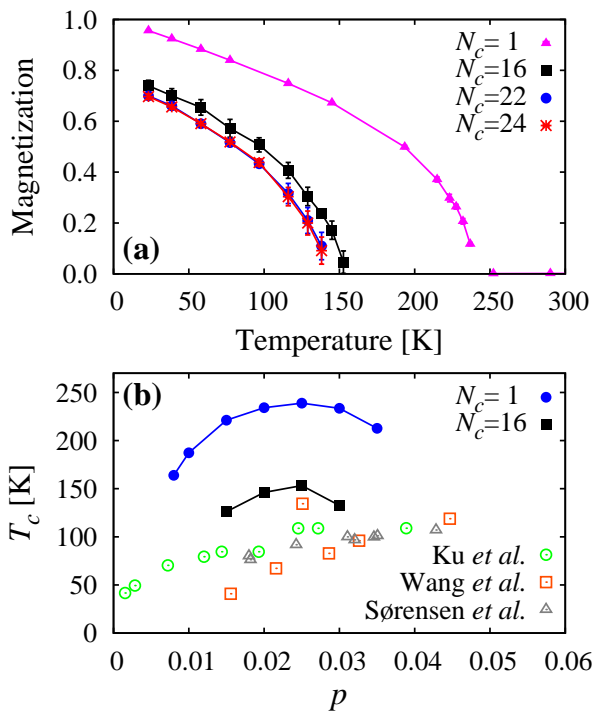}
\caption{\label{fig:1} (Color online) (a) Magnetization of Ga$_{1-x}$Mn$_x$As
calculated by DMFT ($N_c=1$) and DCA ($N_c=16$, $N_c=22$, and $N_c=24$)
with Mn doping $x=0.05$ and hole concentration $p=0.025$.
No strain effect is considered.
(b) Ferromagnetic transition temperature ($T_c$) as a function of
hole concentration ($p$) with DMFT ($N_c=1$) and DCA ($N_c=16$)
for $x=0.05$. Experimental results \protect\cite{Sorensen,Ku,Wang04}
 are also shown. The Mn concentration in experiments is
$x$=0.085 (Ku {\it et al.}), $x$=0.017-0.09 (Wang {\it et al.}),
and $x$=0.05 (S{\o}rensen {\it et al.}).}
\end{figure}

Figure~\ref{fig:1}(a) shows the magnetization per Mn ion of
the Ga$_{1-x}$Mn$_{x}$As system
as a function of temperature with DMFT ($N_c=1$) and
DCA ($N_c=16$, $N_c=22$, and $N_c=24$).
We chose three fcc clusters that are \textit{perfect}
according to Betts \textit{et al.} \cite{Betts}.
The difference between DCA and  DMFT stems from
nonlocal effects, not captured in DMFT.
The $T_c$ with DCA is far lower than that obtained with DMFT, approaching
the regime of experimental values \cite{Sorensen,Ku,Wang04}
[see Fig.~\ref{fig:1}(b)].
Another important point is the reduction of the saturation magnetization
at low temperature, consistent with experiments \cite{Wang04,Esch,Potashnik}.
This behavior is a product of the rotational frustration \cite{Zarand},
to be discussed in detail later.
This effect also reduces $T_c$.
The dependence of $T_c$ on hole concentration ($p$) is shown in Fig.~\ref{fig:1}(b).
$T_c$ attains a maximum value when hole concentration is half of Mn concentration,
consistent with previous DMFT studies \cite{DMS_DMFT}.

\begin{figure}[tbp]
\includegraphics[width=8.2cm]{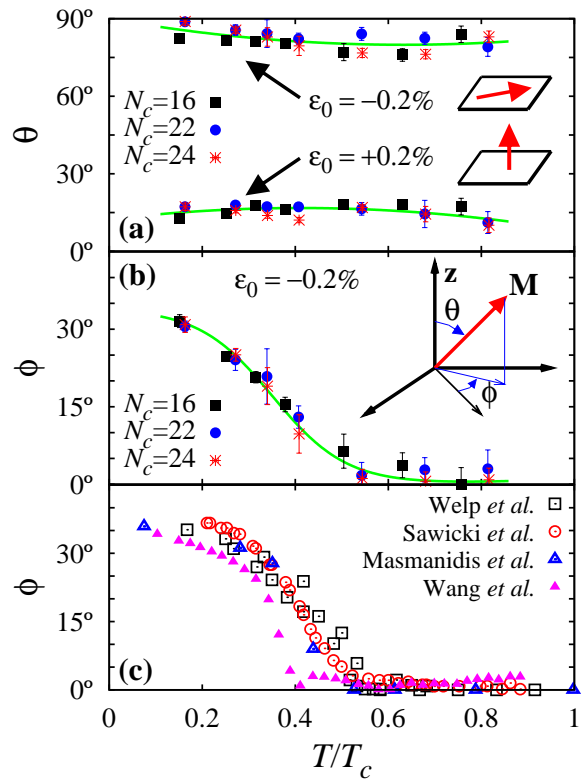}
\caption{\label{fig:2} (Color online) (a) Polar angle ($\theta$) of the magnetization 
as a function of the normalized temperature ($T/T_c$) for two strain values.
Compressive ($\varepsilon_0 = -0.2\%$) and tensile ($\varepsilon_0 = +0.2\%$) strain 
induce in-plane and perpendicular magnetic anisotropy, respectively.
(b) Azimuthal angle ($\phi$) of the magnetization with respect to
the [110] direction with compressive strain.
Experimental results \cite{Sawicki,Wang,Welp,Masmanidis} are provided
in (c) to compare with (b).}
\end{figure}

Next, we studied the magnetic anisotropy of Ga$_{1-x}$Mn$_{x}$As.
The magnetic anisotropy of this system depends on strain, hole concentration,
and temperature in a complicated manner, but generally it has in-plane 
anisotropy with compressive strain and perpendicular-to-plane
anisotropy with tensile strain \cite{Ohno98}.
With compressive strain, the magnetization changes direction
within plane from [110] or [1\={1}0] at high temperature to [100] or [010]
at low temperature \cite{Sawicki,Wang,Welp,Masmanidis} [see Fig.~\ref{fig:2}(c)].
As is shown in Fig.~\ref{fig:2}(a) and \ref{fig:2}(b), the DCA reproduces experimental results
on the dependence of magnetic anisotropy on strain and temperature
remarkably well.

\begin{figure}[tbp]
\includegraphics[width=8.2cm]{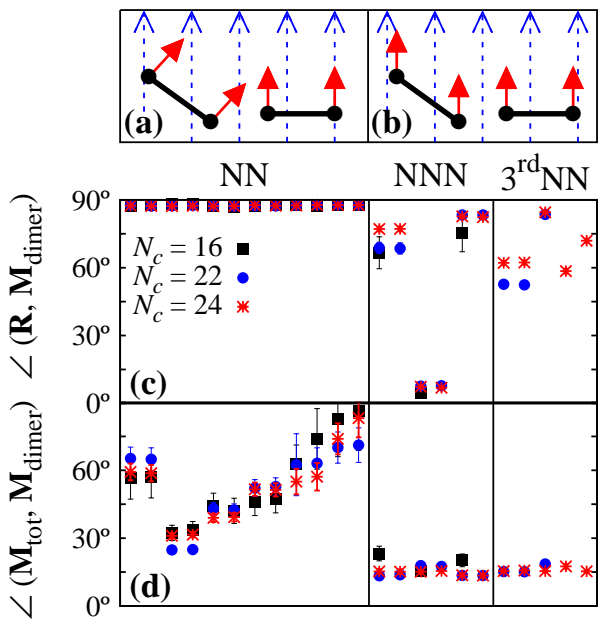}
\caption{\label{fig:3} (Color online)
(a)     DCA results at low temperature show that the dimer magnetization
($\mathbf{M}_{\rm dimer}$) is prefered to be perpendicular to the vector connecting 
the two Mn ions ($\mathbf{R}$) for nearest-neighbor (NN) Mn-dimers.
(b) For larger dimers, $\mathbf{M}_{\rm dimer}$ aligns with the total magnetization
($\mathbf{M}_{\rm tot}$) irrespective of $\mathbf{R}$.  The dotted and solid arrows represent
$\mathbf{M}_{\rm tot}$ and $\mathbf{M}_{\rm dimer}$, respectively.
(c) and (d) Angle between $\mathbf{M}_{\rm dimer}$ and $\mathbf{R}$ and 
between $\mathbf{M}_{\rm dimer}$ and $\mathbf{M}_{\rm tot}$, respectively,
at $T$=23.2 K, $x$=0.05, $p$=0.025, and $\varepsilon_0$=$-$0.2\%.
The left panel is for the 12 NN dimers
[\textit{e.g.}, when the two
Mn ions are at $(0, 0, 0)$ and $(a/2, a/2, 0)$],
the middle panel is for the 6 next-nearest-neighbor (NNN) dimers
[\textit{e.g.}, when the two Mn ions are at $(0, 0, 0)$ and $(a, 0, 0)$],
and the right panel is for the third-nearest-neighbor (3$^{\rm rd}$NN) dimers
[\textit{e.g.}, when the two Mn ions are at $(0, 0, 0)$ and $(a, a/2, a/2)$].
}
\end{figure}

While the strain dependence of the magnetic anisotropy was explained within
the mean-field theory \cite{Abolfath,Dietl01}, the spin reorientation
within the plane has not been explained yet.
In the absence of strain, Ga$_{1-x}$Mn$_{x}$As has diagonal magnetic anisotropy
within the mean-field theory because the heavy holes, which dominate at low carrier density,
have larger density along the diagonal direction in $\mathbf{k}$-space \cite{Abolfath}.
With compressive strain, it has the anisotropy in [110] or [1\={1}0].
This explains the magnetic anisotropy at high temperature.

\begin{figure}[tbp]
\includegraphics[width=8.2cm]{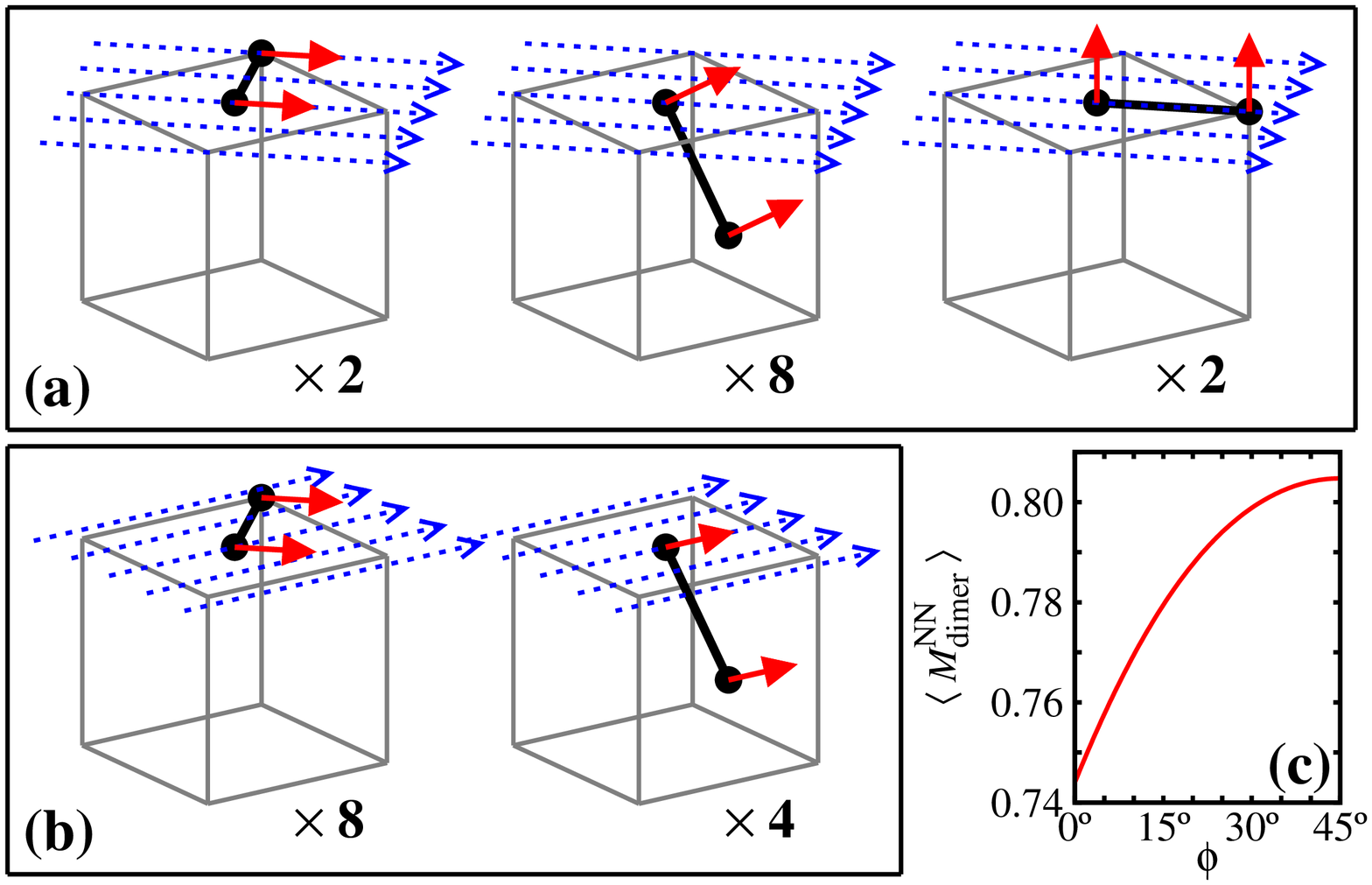}
\caption{\label{fig:4} (Color online) 
(a) Magnetic configurations of the 12 nearest-neighbor (NN) Mn-dimers
($\mathbf{M}_{\mathrm{dimer}}^{\mathrm{NN}}$) that maximize the total 
magnetization ($\mathbf{M}_{\mathrm{tot}}$) when $\phi=0^{\circ}$ 
(\textit{i.e.} $\mathbf{M}_{\mathrm{tot}}$ is along [110] or [1\={1}0]).
The dotted and solid arrows represent
$\mathbf{M}_{\mathrm{tot}}$ and $\mathbf{M}_{\rm dimer}^{\mathrm{NN}}$, respectively.  The numbers below each diagram indicate the degeneracy of 
the Mn-dimer configurations. 
The dimer configuration with $\mathbf{M}_{\mathrm{dimer}}^{\mathrm{NN}}$ 
perpendicular to the vector connecting the two Mn ions ($\mathbf{R}$)
is energetically favored.  
(b) Same as (a) but when $\phi=45^{\circ}$ (\textit{i.e.} $\mathbf{M}_{\mathrm{tot}}$ is along [100] or [010]).
(c) Maximum value of the average $\mathbf{M}_{\mathrm{dimer}}^{\mathrm{NN}}$
vs.\ $\phi$ .
}
\end{figure}

The [100] or [010] anisotropy at low temperature is due
to the cluster anisotropy originating from the anisotropic
interaction between neighboring Mn ions.
Because Mn ions are distributed randomly throughout the system, the
number of Mn ions within a cluster varies between zero and $N_c$.
We call a cluster a monomer (dimer) when it includes one Mn ion (two Mn ions).
All possible distributions of Mn ions are considered effectively
in this calculation, but at low doping, the magnetic properties are dominated
by Mn-monomers and Mn-dimers. Since there is no cluster anisotropy
in monomers, we investigate magnetization of Mn-dimers in detail.
Because of translational symmetry, we need to consider only $N_c-1$ dimers.
The two Mn ions are nearest-neighbors in 12 dimers, next-nearest-neighbors
in 6 dimers, and third-nearest-neighbors in $N_c-19$ dimers.
Figure~\ref{fig:3} shows the magnetization direction of each dimer obtained by
the Monte-Carlo method at low temperature.
The magnetization of the nearest-neighbor Mn-dimer
is always perpendicular to the vector connecting the two Mn ions.
This cluster anisotropy prevents the magnetic moment of some dimers 
from aligning parallel to the total magnetization and leads to the 
rotational frustration \cite{Zarand,Moreno06}.
When the two Mn ions are farther apart, this anisotropy is very weak
and magnetization of the dimer aligns parallel to the total magnetization.

The cluster anisotropy can also explain
the enhancement of $T_c$ up to 260 K in the quasi-two-dimensional
$\delta$-doped systems \cite{Nazmul}.
When the Mn ions are within one plane, all the nearest-neighbor Mn-dimers
can point in the same direction
(perpendicular-to-plane direction) without rotational frustration.
This may induce a larger saturation magnetization and higher $T_c$ in these
systems.

Notably, due to this cluster anisotropy, the maximum total magnetization 
depends on the magnetization direction,
and this dependence introduces 
another type of magnetic anisotropy. When we assume the magnetic moment is
perpendicular to the vector connecting
the two Mn ions and in-plane magnetization,
the maximum value of the average magnetization of the 12 nearest-neighbor dimers
is calculated to be
\begin{eqnarray}
\langle M_{\mathrm{dimer}}^{\mathrm{NN}}\rangle &=&
\frac{1}{6} \left[ \cos\phi + \sin\phi 
 + \sqrt{3+\sin(2\phi)} \right. \nonumber \\
 && ~~ \left. + \sqrt{3-\sin(2\phi)} \right]
\label{Eq:M_dimer}
\end{eqnarray}
for the fcc lattice. This becomes maximal
when the magnetization is along [100] or [010],
as shown in Fig.~\ref{fig:4}(c).
Since larger magnetization leads to lower magnetic energy,
it introduces magnetic anisotropy along [100] or [010].
We note that this effect becomes unimportant at high
temperature, where the total magnetization is small.
Thus, the spin reorientation from [110] or [1\={1}0]
at high temperature to [100] or [010] at low temperature
is captured within our calculation. 
This behavior arises from the multi-impurity scattering and
cannot be obtained within the mean-field theory or the DMFT.

In summary, we investigated the magnetic properties of the prototypical
DMS system Ga$_{1-x}$Mn$_{x}$As by DCA together with 
the $\mathbf{k}\cdot\mathbf{p}$ method.
We showed that nonlocal effects, not included in the mean-field 
theory or the DMFT but included in the DCA for $N_c>1$, are very important 
to quantitatively explain the $T_c$, the saturation magnetization,
and the magnetic anisotropy of this material.
We find that the edge-direction anisotropy at low temperature
is due to the cluster anisotropy, and this allows us to reproduce
the spin reorientation in this system remarkably well.

We acknowledge helpful discussions with Z. Xu.
This work was supported by the National Science Foundation through
OISE-0730290, DMR-0548011, and DMR-0706379.
BM acknowledges support from the U.S. Department of Energy,
Office of Basic Energy Sciences, under contract DE-AC02-76SF00515.
Portions of this research were conducted with high performance
computational resources provided by the Louisiana Optical
Network Initiative (http://www.loni.org).

\end{document}